\documentstyle[aps,psfig,epsfig,rotating,preprint,prb]{revtex}

\begin{document}

\author{ A.V. Zvelindovsky \and G.J.A. Sevink \and J.G.E.M. Fraaije}

\address{Faculty of Mathematics and Natural Sciences, University of
Groningen, Nijenborgh 4, 9747 AG Groningen;\\ Leiden Institute of
Chemistry, Leiden University, Einsteinweg 55, 2300 RA Leiden, The
Netherlands}

\title{Shear-induced Transitions in Ternary Polymeric System}

\maketitle

\begin{abstract}
The first three-dimensional simulation of shear-induced phase
 transitions in a polymeric system has been performed. The method is
 based on dynamic density-functional theory.  The pathways between a
 bicontinuous phase with developing gyroid mesostructure and a
 lamellar/cylinder phase coexistence are investigated for a mixture of
 flexible triblock ABA-copolymer and solvent under simple steady
 shear.
\end{abstract}

Various self assembly systems such as lyotropic liquid crystals,
 surfactants, block copolymers can form ordered mesophases (lamellar,
 cylindrical, spherical, etc). These phases have received much
 attention because of the fundamental interest to establish universal
 laws for self-organization phenomena, and also because of the wide
 range of applications in materials science
 \cite{THE,BaFr,matS,doi,po,pio}.

An interesting issue in the design of new materials is modulation of
phase behaviour by external and internal factors such as flows
\cite{science}, reactions \cite{last_nat}, temperature inhomogeneity
\cite{Hash}, confinements and surfaces \cite{sev}. In particular,
externally applied shear flows are found to lead to macroscale order
in block copolymer systems. Moreover, shear introduces a new kind of
phase behaviour of block copolymer systems, the so-called
orientational phase transitions
\cite{fredrickson94,wiesner97,tepe,morozov2000}.

So far, lamellar, hexagonal cylindrical and cubic micellar phases of
block copolymers under shear have been thoroughly investigated in
experiments \cite{THE,science,wiesner97,tepe,schmidt98}.  Theoretical
understanding is partly reached for the lamellar geometry
\cite{fredrickson94,cates89,goulian95,Dro} and for the hexagonal
cylindrical phase \cite{morozov2000,marques,mut}.  Experimental
observation of a sponge phase in low molecular weight surfactants
under shear has been carried out recently in \cite{Mah98}. Elongation
of mesostructures has been recently found in a sheared isotropic
bicontinuous polymer phase \cite{Qiu98}.  More complex phases such as
gyroid and coexistence of phases in shear flow still require
experimental examination.

Previously, computer simulations of polymer morphologies under shear
have been carried out for lamellar and cylindrical phases in 2D
systems (see references in \cite{THE}).  Recently we have reported 3D
density functional calculations for these two phases
\cite{zvPRE,zvCP}.  Here we report on the effect of shear on more
complex phases --- bicontinuous gyroid and lamellar/cylinder
coexistence in a three component copolymer system. We have observed
shear-induced transitions between these states.  Many shear-induced
transitions are found in experiments and theory \cite{THE}.
Transitions we present here require future experimental and
theoretical investigation.

The time evolution of the density field under simple steady shear
flow, $v_x=\dot\gamma y,$ $v_y=v_z=0,$ can be described by a time
dependent Landau-Ginzburg type equation with a convective term
\cite{doi,zvPRE,onuki97c} and a stochastic term
\cite{fraaije97a,macro}.  In general, the velocity field can be found
from hydrodynamics. For a system with different viscosities of the
components, this leads to an adaption of the linear velocity profile
\cite{fredrickson94}.  Such refinement is important for the
description of temperature dependency of the orientation of
mesostructure lattice in the gradient-vorticity plane at high shear
(see discussion in \cite{morozov2000}).  However, in the weak
segregation regime, when there are no steep concentration gradients,
the basic features of the process of alignment in flow (regardless of
fine details of orientational transitions) can be described accurately
by diffusion-convection equation, with an imposed velocity profile
\cite{Dro,zvPRE,zvCP}.

 In contrast to traditional schemes of polymer phase separation
dynamics where a Landau Hamiltonian is used with vertex functions
calculated following the Random Phase Approximation (see {\it e.g.}
\cite{THE}), we numerically calculate the ``exact'' free energy $F$ of
polymer system consisting of Gaussian chains in mean field environment
using path integral formalism \cite{fraaije97a,macro}.  The benefit of
our approach is that it avoids the truncation of the free energy and
therefore represents intermediate (metastable) states more accurately.

 Some time ago, Matsen and Schick \cite{matS} introduced a powerful
 method for SCF calculations of equilibrium block copolymer
 morphologies.  Our approach uses essentially the same free energy
 functional and complements the static SCF calculations by providing a
 dynamical picture of the system which is crucial for systems under
 shear \cite{doi,onuki97c,ka}.  In our method the calculations are in
 direct space, without any bias with respect to the morphology
 formation.

 The system presented here is a mixture of Gaussian chains $E_3P_9E_3$
(bead names are arbitrary) and solvent which is parameterized to model
60\% aqueous solution of a triblock copolymer Pluronic surfactant PL64
$(EO)_{13}(PO)_{30}(EO)_{13}$ with the hydrophobic block in the middle
\cite{macro}.  The choice is justified by huge variety of available
experimental data.  In experimental phase diagram, 60\% polymer
concentration corresponds to a very complex phase coexistence region
\cite{Al98Lang}.

Fig. 1 gives a schematic overview of the simulation of the system in a
3D box ($64 \times 64 \times 64 $). The starting configuration is a
homogeneous distribution of the components.  The system
(Fig. \ref{fig1}) demonstrates the development of a bicontinuous
morphology with clear gyroid-type connectivity (Fig. \ref{fig2}),
however still without global symmetry throughout the sample. The
system can remain in this phase very long, slowly rearranging the
structure and keeping gyroid-type connectivity (see also
Fig. \ref{figF}a).

After applying shear ($\widetilde{\dot\gamma}=10^{-3}$) to the
morphology shown in Fig. \ref{fig1}, the system slowly
deforms. However, the connectivity hardly changes: the system remains
bicontinuous and similar to the morphology shown in Fig. \ref{fig1}
even at large shear strains $\gamma > 1 (100\%)$. While shearing
continues, connections in the system start to break, separate pieces
of the structure reconnect to form a pattern aligned in the flow
direction.  New structure consists of coexisting lamellae and
hexagonally packed cylindrical clusters, as shown in Figs. \ref{fig3}a
and \ref{fig4}a.

\begin{figure}
\setlength{\unitlength}{1cm}
\begin{picture}(10,2)
\put(-0.6,0){\epsfig{file=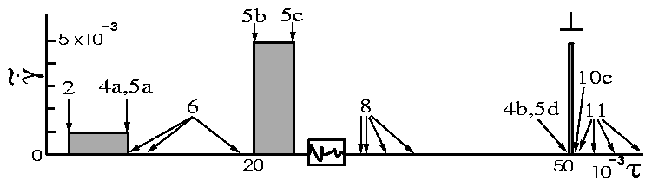,width=8.5cm}}
\end{picture}
\caption{The sample history. Arrows indicate figures with
snapshots. The box with the curve indicates a high noise region and
the symbol $\perp$ indicates change in shear direction.
$\widetilde{\dot\gamma}=\Delta t \dot\gamma$ is the dimensionless
shear rate and $\tau$ is dimensionless time with time step
$\Delta\tau=\beta^{-1} M h^{-2} \Delta t = 0.5$ (see \cite{macro}).  }
\label{fig2a}
\end{figure}

\begin{figure}
\setlength{\unitlength}{1cm}
\begin{picture}(9,7)
\put(0.5,0){\psfig{file=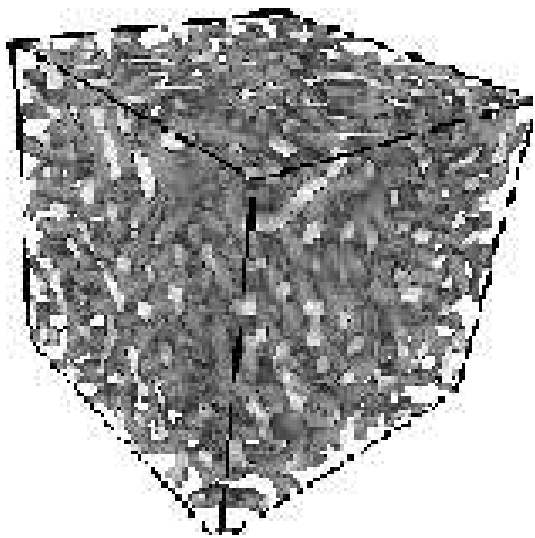,height=7cm}}
\end{picture}
\caption{An isosurface of $PO$ blocks at $\tau=2000$ (no shear).}
\label{fig1}
\end{figure}

\begin{figure}
\setlength{\unitlength}{1cm}
\begin{picture}(10,2.5)
\put(0.2,0.5){\epsfig{file=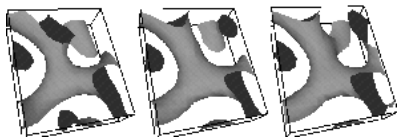}}
\end{picture}
\caption{Evolution of a detail of the bicontinuous morphology in the
absence of shear at $\tau=2000$, $4975$, $7500$ (from left to right).}
\label{fig2}
\end{figure}

\begin{figure}
\setlength{\unitlength}{1cm}
\begin{picture}(10,11)
\put(-0.3,0){\epsfig{file=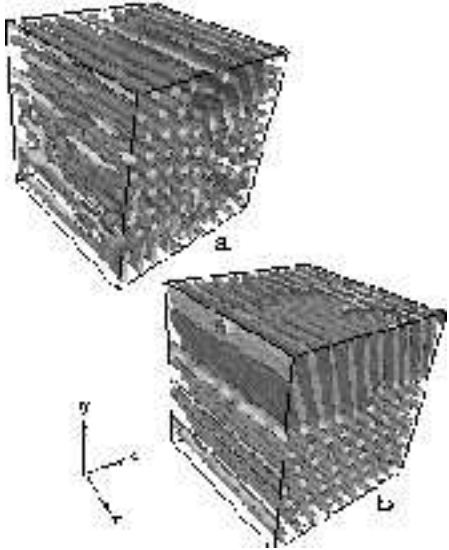}}
\end{picture}
\caption{Morphology of $PO$ blocks at $\tau=7500$ (a) and 50000 (b).}
\label{fig3}
\end{figure}

\begin{figure}
\setlength{\unitlength}{1cm}
\begin{picture}(10,6.5)
\put(0.5,0.5){\epsfig{file=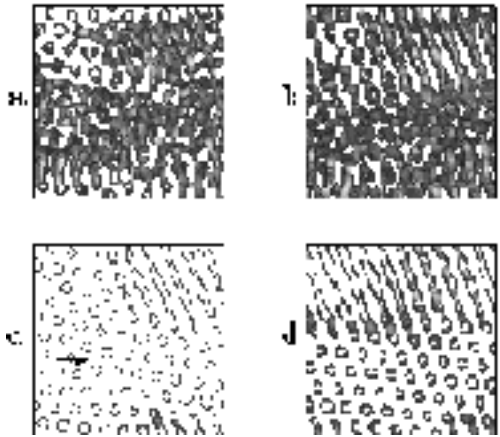}}
\end{picture}
\caption{Orthogonal projection in x-direction of the morphologies at
different times: $\tau=7500$ (a), 20000 (b), 23500 (c), 50000 (d).
The arrow in (c) denotes a dislocation.}
\label{fig4}
\end{figure}

\begin{figure}
\setlength{\unitlength}{1cm}
\begin{picture}(10,2.5)
\put(0.2,0){\epsfig{file=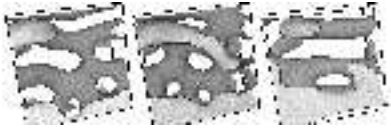}}
\end{picture}
\caption{Detail of a perforated lamellae transformation: $\tau$=7500,
9500, 18500 (from left to right).}
\label{fig6}
\end{figure}

\begin{figure}
\setlength{\unitlength}{1cm}
\begin{picture}(10,4)
\put(-0.2,0){\epsfig{file=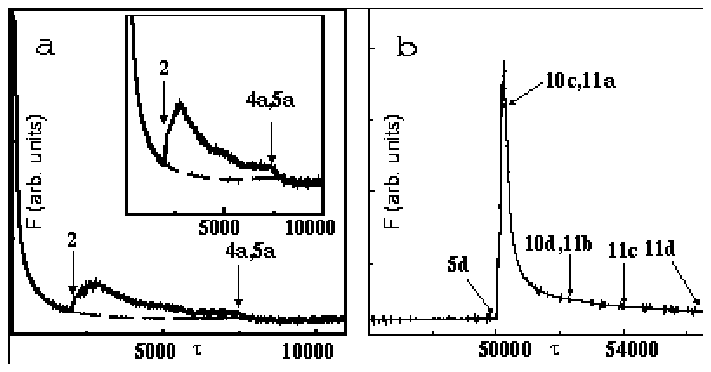,width=8cm}}
\end{picture}
\caption{Free energy as function of the time for the first (a) and
 the last (b) period of shear. The inset is magnified
 representation of the left plot.  Arrows indicate figures with
 snapshots. First and second arrows from the left correspond to
 starting end stopping times of shearing ({\it cf.}
 Fig.\ref{fig2a}). The dashed line on the left graph corresponds to
 bicontinuous phase evolution in the absence of shear (see
 Fig.\ref{fig2}).  }
\label{figF}
\end{figure}

Stopping the shear at some point ($\tau=7500$ ($\gamma =11$),
Fig. \ref{fig2a}) leads to the reorganization of the structure via
migration of defects (holes in lamellae, necks between cylinders),
Fig. \ref{fig6}.  The time evolution of deformed and broken
bicontinuous structure involves a lot of perforated lamellar
clusters. Migration of defects in these clusters leads sometimes to
intermediate structures with locally hexagonal-like arrangement of
holes (Fig. \ref{fig6}, middle picture). After long relaxation without
shear the structure remains a coexistence of defected cylinders and
lamellae (Fig. \ref{fig4}b) but with a better ordering compared to the
moment of stopping shear (Fig. \ref{fig4}a). The free energy plot
(Fig. \ref{figF}) provides information which state is
metastable. Bicontinuous and coexistence states are very close in free
energy value in plateau region (Fig. \ref{figF}a), but the last
is slightly lower.

\begin{figure}
\setlength{\unitlength}{1cm}
\begin{picture}(10,2.5)
\put(0.5,0){\epsfig{file=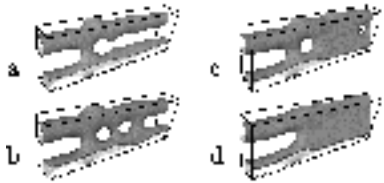}}
\end{picture}
\caption{Detail of neck dynamics: $\tau=30000$ (a), 30500 (b), 32500
(c), 35000 (d).}
\label{fig5}
\end{figure}

\begin{figure}
\setlength{\unitlength}{1cm}
\begin{picture}(10,2)
\put(0,0){\epsfig{file=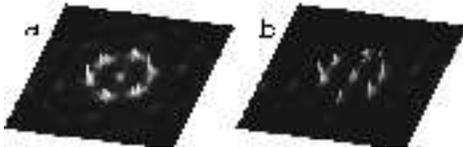}}
\end{picture}
\caption{ 3D structure factor summed in x-direction (log-scale) for
 $\tau=23500$ (a), ({\it cf.}  Fig.5c), and $\tau=50000$ (b),
 ({\it cf.} Fig.5d).}
\label{fig7}
\end{figure}

A second period of stronger shear ($\widetilde{\dot\gamma}=5 \cdot
10^{-3}$) very fast breaks up the remaining connections and the system
flows as a lamellar/cylinder coexistence without changing
lamellar/cylinder volume ratio while shear continues, $\tau=23500$
($\gamma = 35$), Fig. \ref{fig4}c. The lamellae have only a few holes.
The cylinder region no longer has "neck" defects and the only
remaining defect is a long living dislocation which remains stable
during the whole period of shearing. As a result the hexagonally
packed cylinders consist of two clusters with different internal
hexagonal orientation (as can also be seen from the double peaks in
the structure factor, Fig. \ref{fig7}a).  After stopping shear at
$\gamma =35$ (Fig. \ref{fig4}c) the system relaxes to a
lamellar/cylinder coexistence in which the hexagonal lattice has only
one orientation (Figs. \ref{fig3}b, \ref{fig4}d and
\ref{fig7}b). Undulated cylinders partly transform into lamellae via
formation of subsequent necks, Fig. \ref{fig5}. The lamellar phase
still consists of perforated lamellae, with a low fraction of
holes. This coexistence of phases seems to be stable and even
application of a much higher noise ($\Omega=10$ \cite{macro}, the
region is schematically shown in Fig. \ref{fig2a}) does not change the
picture considerably.

Both periods of shear demonstrate that the lamellar cluster becomes
larger after shear is released (Figs. \ref{fig4}a,b and
\ref{fig4}c,d). This is consistent with the fact that switching on the
shear again squeezes the size of the lamellar region,
Figs. \ref{fig4}b and \ref{fig4}c.

\begin{figure}
\setlength{\unitlength}{1cm}
\begin{picture}(10,7)
\put(0.5,0.5){\epsfig{file=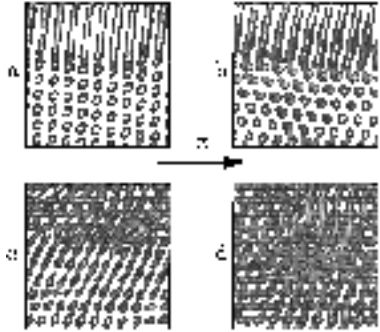}}
\end{picture}
\caption{Results of shear in z-direction, taking the morphology as
 shown in Fig.\ref{fig3}b ( $\tau=50000$), as starting structure.
 Orthogonal projections of the morphologies at $\tau=50050$ (a), the
 end of shear, and $\tau=54050$ (b).  A longer period of shear from
 the same starting structure: $\tau=50300$ (c), the end of shear, and
 at $\tau=52300$ (d).  }
\label{fig8}
\end{figure}

\begin{figure}
\setlength{\unitlength}{1cm}
\begin{picture}(10,2)
\put(-0.3,0.5){\epsfig{file=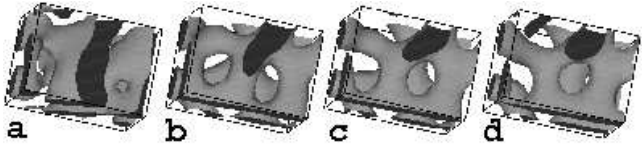,width=8.5cm}}
\end{picture}
\caption{ Detail
 of structure relaxation after shearing in z-direction
at $\tau=50300$ (a)  ({\it cf.} Fig.\ref{fig8}c), $52300$ (b)
 ({\it cf.} Fig.\ref{fig8}d), $54000$ (c), $56500$ (d).
}
\label{fig9}
\end{figure}

The stability of the phase coexistence morphology, Figs. \ref{fig3}b,
 \ref{fig4}d, is challenged by applying shear
 ($\widetilde{\dot\gamma}=5\cdot 10^{-3}$) in the z-direction
 (perpendicular to lamellae/cylinders), Fig. \ref{fig2a}.  In
 Figs. \ref{fig8}a and \ref{fig8}c we see the result of shearing for a
 period of $\tau=50$ ($\gamma = 0.5$) and $\tau=300$ ($\gamma = 3$),
 respectively.  While shearing the lamellae start to tilt and
 thin. The cylinders start to roll over each other and deform to a
 slightly prolate shape in cross-section.  This leads to an
 energetically unfavorable cubic cylindrical lattice,
 Fig. \ref{fig8}a, or to a morphology with very oblong cylinders in
 cross-section and partly broken lamellae,
 Figs. \ref{fig8}c,\ref{fig9}a.  Releasing shear at these different
 stages leads to different phenomena. After a relatively small
 distortion of the system ($\gamma = 0.5$, Fig. \ref{fig8}a) in short
 time the system relaxes back from cubic to hexagonal packed
 cylinders, keeping the lamellar cluster intact,
 Fig. \ref{fig8}b. However, if shear is stopped after a larger
 distortion ($\gamma = 3$, Fig. \ref{fig8}c), then very fast necks
 form throughout the sample, Fig.\ref{fig8}d. The structure becomes
 bicontinuous with sometimes obviously gyroid-like connectivity,
 Fig. \ref{fig9}c.  However the global structure is different from the
 initial morphology in Fig.\ref{fig1}. The process of relaxation after
 stopping shear goes via following stages: frustrated
 lamellae/cylinders $\rightarrow$ bicontinuous phase $\rightarrow$
 lamellae/cylinders.  The first stage is relatively fast whereas the
 last is slow, which can be viewed in hardly noticeable changes in
 structure in Figs.\ref{fig9}c and d.  Free energy plot
 (Fig. \ref{figF}b) also demonstrates that system tends to relax
 back to a coexistence state. Thus, in this case the bicontinuous
 phase is a long living intermediate stage with the free energy very
 close to lamellae/cylinders coexistence state.

In summary, we have performed the first 3D shear simulation of a
bicontinuous ABA-copolymer/solvent system.  New shear induced phase
transitions from bicontinuous phase to lamellae/cylinder coexistence
and back have been detected. These two states have very close free
energy values and can be separated by a barrier in the free energy
landscape. As a result the polymeric system can be trapped in either
of them.

 A.V.Z.  and G.J.A.S. acknowledge support of the MesoDyn project
ESPRIT No. EP22685 of the European Community. We acknowledge support
of NCF (Stichting Nationale Computer Faciliteiten).

\end{document}